\documentstyle[aps,twocolumn,graphicx]{revtex}

\begin{document}
\draft

\title{Photon Correlation Spectroscopy of a Single Quantum Dot}

\author{A.~Kiraz,$^1$ S.~F\"alth,$^1$ C.~Becher,$^{1,2}$ B.~Gayral,$^1$
W.V.~Schoenfeld,$^3$ P.M.~Petroff,$^3$ Lidong~Zhang,$^1$ E.~Hu,$^1$ and
A.~Imamo\u{g}lu,$^{1,4}$}
\address{$^1$ Department of Electrical and Computer Engineering, University
of California, Santa Barbara, California 93106}
\address{$^2$ Institut f\"ur Experimentalphysik, Universit\"at Innsbruck,
A-6020 Innsbruck, Austria}
\address{$^3$ Materials Department, University of California, Santa Barbara,
 California 93106}
\address{$^4$ Department of Physics, University of California, Santa Barbara,
 California 93106}

\twocolumn[\hsize\textwidth\columnwidth\hsize\csname @twocolumnfalse\endcsname

\maketitle

\date{\today}


\begin{abstract}

We report photon correlation measurements that allow us to observe unique signatures of
biexcitons in a single self-assembled InAs quantum dot. Photon correlation measurements of
biexciton emission exhibit both bunching and antibunching under continuous-wave excitation
while only antibunching is observed under pulsed excitation. Cross-correlation between
biexciton and single-exciton peaks reveal highly asymmetric features, demonstrating that
biexciton and exciton emissions have strong correlations due to cascaded emission. The
anticipated correlation between the polarization of exciton and biexciton emissions
however, is absent under our excitation conditions. Photon correlation measurements also
provide evidence for the identification of the charged exciton emission.

\end{abstract}



\vskip2pc] \narrowtext


It is by now widely accepted that various quantum dot (QD) structures exhibit features in
transport \cite{KasPT93,DrePRL94} or optical spectroscopy
\cite{GamSci96,ZrePRL94,DekPRL98} that indicate full three dimensional confinement of
carriers. Identification of QDs as artificial atoms has been strengthened by the recent
observation of strong photon antibunching in single-exciton emission
\cite{MicNat00,BecPRB01}, which is the typical signature of an anharmonic quantum system:
after a photon is emitted from a single two-level (anharmonic) emitter, the system is
necessarily in the radiatively inactive ground state and a second photon cannot be emitted
immediately after the first one. Even though the coherence properties of QD single-exciton
emission closely follow those of atoms, the overall spectral features of single QDs are
significantly more complicated. Since the size of QDs is roughly two-orders of magnitude
larger than those of atoms, multi-particle excitations give rise to emission peaks with
energies comparable to that of a single-exciton. Of primary importance in QD spectroscopy
is the biexciton state, which corresponds to a doubly-excited QD with completely filled
lowest electron and hole energy levels. When the biexciton state decays by radiative
recombination, the final-state is a single-exciton state and the generated photon is
shifted as compared to the single-exciton emission due to Coulomb interaction between the
carriers. Biexciton emission in QD spectroscopy has been traditionally identified using
the (quadratic) pump-power dependence of the corresponding peak.

In this letter, we demonstrate that photon correlation measurements provide a powerful
tool for characterizing the multiexciton spectral features of QDs. Our measurements
provide a strong support for the identification of a biexciton emission peak, by
demonstrating its strong correlations with the subsequent single-exciton emission. We
observe that biexciton intensity autocorrelation exhibits bunching together with
antibunching or only antibunching under continuous-wave (cw) excitation depending on the
excitation level. In contrast, we find strong antibunching under pulsed excitation. The
large difference between the levels of antibunching under continuous-wave and pulsed
excitations points out to the importance of excitation mechanism and the role of free
carriers in QD physics. The lack of polarization correlation between biexciton and
single-exciton emissions indicates that spin dephasing is likely to play a key role under
non-resonant excitation. We also observe that a third emission peak in QD spectra exhibits
strong correlations with both exciton and biexciton fluorescence: we argue that these
correlation signatures suggest the identification of this additional line as a charged
exciton emission.

Our self-assembled InAs QDs were grown by molecular beam epitaxy (MBE) using the partially
covered island technique \cite{GarAPL98}. Growth resulted in typically lens shaped QDs
with a base diameter of 40-50~nm and a height of 3~nm, having their single-excitonic
emissions between 925~nm and 975~nm in the spectrum. In our sample, the QDs were embedded
in the center of a 200~nm thick GaAs microdisk structure located above a 0.5~$\mu$m thick
Al$_{0.65}$Ga$_{0.35}$As post. The diameter of the disks was 5~$\mu$m and the average
number of QDs within the disks was less than 1. Details of the microdisk processing can be
found elsewhere \cite{MichAPL00}. Our experimental setup consisted of a combination of a
low-temperature diffraction-limited scanning optical microscope and a Hanbury Brown and
Twiss (HBT) \cite{HanNat56} setup for photon correlation measurements. The QD sample was
mounted in a He-flow cryostat and cooled to 4-7~K. The cryostat was moved by a
computer-controlled translation stage, thus allowing for scanning across the sample. The
QDs were optically excited either with a continuous wave diode laser (operating at
785~nm), a continuous wave Ti:Sa laser (operating at 760~nm) or a mode-locked Ti:Sa laser
(82~MHz, 250~fs, operating at 790~nm), generating electron-hole pairs in the GaAs barrier
layer which are subsequently captured by the QDs within a short timescale ($<$ 35~ps)
\cite{RayPRB96}. The same microscope objective (NA = 0.55) mounted outside the cryostat
was used to both excite the QDs (diffraction limit : $\sim1.7~\mu$m) and collect the
emitted fluorescence light. The collected light was spectrally filtered either by a 0.5~m
monochromator or narrowband interference filters (FWHM=0.5 or 1~nm) before being detected
in the HBT setup, consisting of a 50/50 beamsplitter and two single-photon-counting
avalanche photodiodes. The APDs were connected to the start and stop inputs of a time to
amplitude converter (TAC). The TAC output was stored in a multichannel analyzer (MCA) to
yield the number of photon pairs $n(\tau)$ with arrival time separation of
$\tau=t_{start}-t_{stop}$. An electronic delay was introduced into the stop channel in
order to measure the photon correlation for negative time delays. The time resolution of
the HBT setup was 420~ps. Under our experimental conditions with detection efficiency of
about $0.1 \%$, the measured photon-pair distribution directly yields the normalized
(second-order) intensity auto-correlation (coherence) function $g^{(2)}(\tau)=\langle :
I(t)I(t+\tau) :\rangle / \langle I(t) \rangle^2$, after proper normalization of the MCA
output. Here $ : : $ denotes normal ordering and $I(t)$ is the intensity operator.

\begin{figure}
\centerline{\scalebox{0.5}{\includegraphics{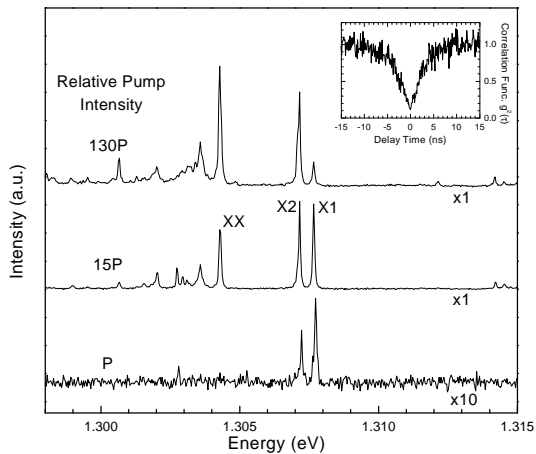}}} \caption{PL spectra for the
quantum dot studied in this letter for various powers of the cw diode laser at 785~nm.
Indicated are the three peaks that we focus on. With increasing powers, the dominant line
is successively X1, then X2, then XX. All lines saturate and then decrease with increasing
power. Inset: Photon correlation measurement carried out using the X1 emission, showing
strong antibunching.}
\end{figure}

Fig.~1 shows the power dependent photoluminescence (PL) spectra of the single QD that we
analyze. At low pump powers, the single-exciton emission peak (X1) dominates the spectrum.
At higher pump powers, we observe that two other peaks become dominant: among these, the
lower energy one (XX) has an energy (red) shift of 3.5~meV from X1 and its intensity has a
quadratic dependence on pump power; these are typical signatures for biexciton emission in
self-assembled InAs QDs. The third peak (X2) is red-shifted from the single-exciton peak
by about 500~$\mu$eV. All three emission peaks are resolution limited at 70$~\mu$eV, and
none of them is polarized. To ensure that X1 originates from a single QD exciton emission,
we have carried out photon auto-correlation measurements where both APDs were illuminated
by X1 emission : under cw and pulsed excitation, X1 emission was found to exhibit perfect
antibunching (Fig.~1 inset) and single-photon source operation \cite{MichSci00,SenPRL01},
respectively. We have also performed time correlated single photon counting experiments on
X1, X2, and XX emissions to measure their lifetimes \cite{TCSPC}. Those measurements were
performed in the very low excitation regime where the decay times of the resulting spectra
were determined by the lifetimes of the corresponding emissions \cite{BacPRL99}. From the
measured spectra we deduced lifetimes of 3.6~ns, 3.7~ns and 2.6~ns for the X1, X2, and XX
emissions respectively. The resultant ratio of $\tau_{X1}/\tau_{XX}=1.4$ is consistent
with exciton and biexciton lifetime measurements performed on CdSe/ZnSe QDs
\cite{BacPRL99}. We also note that, due to the different photonic environment created by
the microdisk, the single-exciton lifetime we observe (3.6~ns) is larger than typical
lifetimes measured ($\sim$1~ns) in bulk QD samples \cite{BecPRB01}.

\begin{figure} \centerline{\scalebox{0.5}{\includegraphics{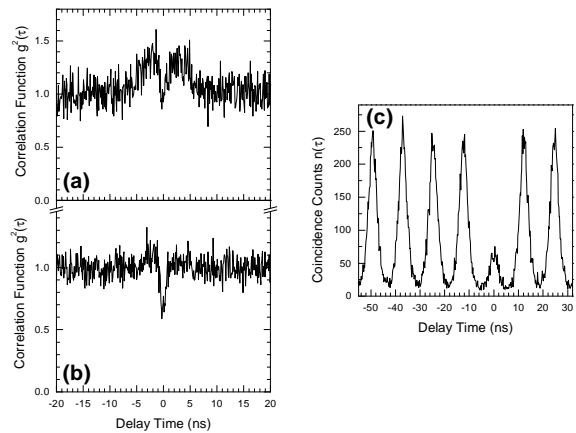}}}
\caption{Measured second order correlation function of the XX line. Under cw-excitation
(Ti:Sa laser, 760~nm) (a) an antibunching dip ($g^{(2)}=$0.95) together with a bunching
peak ($g^{(2)}=$1.4), and (b) an antibunching dip ($g^{(2)}=$0.6) are visible at X1
emission intensities 0.7 and 1.0 of the (exciton) saturation level, respectively. (c)
Under pulsed excitation (82~MHz, 250~fs, at 790~nm), the relative area of the central peak
compared to the other peaks is 0.3, indicating strong antibunching in contrast with the cw
results.}
\end{figure}

Since the QD spectrum is anharmonic, it could be argued that the measurement of
$g^{(2)}(\tau)$ for the XX line will also exhibit simple antibunching, as has been
observed for single-exciton emission. Figures~2(a) and 2(b) show $g^{(2)}(\tau)$ for the
XX emission under cw excitation, at pump powers corresponding to X1 emission intensities
that are 0.7 and 1.0 of the (exciton) saturation level, respectively. Both curves,
obtained using a 0.5~nm interference filter, exhibit antibunching ($g^{(2)}(0)=0.95$ in
Fig.~2(a), $g^{(2)}(0)=0.6$ in Fig.~2(b)) with similar decay times of 1~ns. The curve in
Fig.~2(a) also exhibits bunching ($g^{(2)}=1.4$) that decays with a decay time of 3.5~ns.
Bunching here originates from the fact that the detection of a photon at the biexciton
transition results in the projection of the QD wave-function onto the single-exciton-state
X1. When the average occupancy of X1 in steady-state is lower than unity,
post-measurement-state has higher occupancy in the single-exciton-state  than
pre-measurement-state, and is more likely to result in re-excitation of the biexciton
state. An analysis of the QD dynamics using 3-level rate equations indicates that
$g^{(2)}(\tau)$ should indeed exhibit bunching that decays in a timescale determined by
the single-exciton lifetime of 3.6~ns which is in agreement with the experimental result
(Fig.~2(a)). This analysis also predicts that antibunching at $\tau=0$ should turn into
bunching in a timescale determined by the biexciton lifetime in the low excitation regime.

We could observe strong bunching but no antibunching in biexciton auto-correlation
measurements when we use a 1~nm interference filter, indicating the importance of
additional broadband radiation at biexciton energy that appears to be correlated with
exciton emission. While the pump laser wavelength has a strong effect on the observability
of biexciton antibunching, pump intensity plays no significant role in determining the
recovery time of the antibunching dip. On the other hand, the stronger biexciton
antibunching under pulsed excitation suggests that the free carriers could still have an
adverse effect (Fig.~2(c)): we remark that under pulsed excitation, free-carriers
recombine in a time-scale that is much faster than the biexciton radiative recombination
time, and therefore their influence on biexciton dynamics is expected to be minimal. The
antibunching dip in Fig.~2(a) is also affected by the fact that due to the presence of
bunching, the correlation function recovers to a value exceeding unity within a biexciton
lifetime; the effect of time resolution is therefore more pronounced in the vicinity of
zero time delay compared to the excitation regime when no bunching is observed
(Fig.~2(b)).

\begin{figure}
\centerline{\scalebox{0.5}{\includegraphics{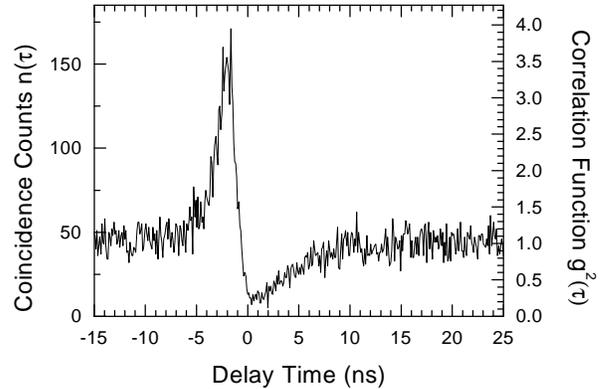}}} \caption{Cross correlation
function between X1 and XX emissions under cw diode laser excitation at 785~nm.  The
signal from X1, filtered by a 0.5~nm interference filter, is sent to the start APD, while
the XX line, filtered by a 1~nm filter, is sent to the stop APD. The positive correlation
($g^{(2)}=$3.4) for $\tau < 0$ followed by the negative correlation ($g^{(2)}=$0.2) for
$\tau>0$ is an evidence for the cascaded emission of photons.}
\end{figure}

Cross-correlation measurements complement the identification of the biexciton emission:
since X1 population is enhanced as a result of the detection of an XX photon, strong
correlations between the X1 and XX emissions can be expected. Fig.~3 shows such a photon
cross-correlation measurement, obtained by illuminating the start APD by the X1 emission
and stop-APD by the XX emission. The depicted quantity here is
$\tilde{g}^{(2)}(\tau)=\langle : I_{XX}(t)I_{X1}(t+\tau) : \rangle / (\langle I_{X1}(t)
\rangle \langle I_{XX}(t) \rangle)$, where $I_{X1}(t)$ and $I_{XX}(t)$ are the intensities
of the X1 and XX emissions, respectively. Remarkable features of this cross-correlation
include strong antibunching for $\tau > 0$ and strong bunching for $\tau<0$ with a close
to resolution-limited transition between the two regimes. For $\tau > 0$, suppression of a
joint X1 and XX event arises from the fact that following the detection of an X1 photon,
which projects the QD onto its ground-state, detection of an XX photon is very unlikely.
Strong bunching for $\tau < 0$ follows from the fact that detection of an XX photon
projects the QD onto the X1 state, as discussed earlier. Asymmetry in cross-correlation
measurements under pulsed excitation have been recently reported \cite{ROBERT}. We claim
that the signature depicted in Fig.~3 proves that the XX emission arises from the decay of
the biexciton state into the single-exciton state. The strong antibunching in
cross-correlation is yet another indication that the additional broadband radiation is
correlated with the X1 emission.

\begin{figure}
\centerline{\scalebox{0.5}{\includegraphics{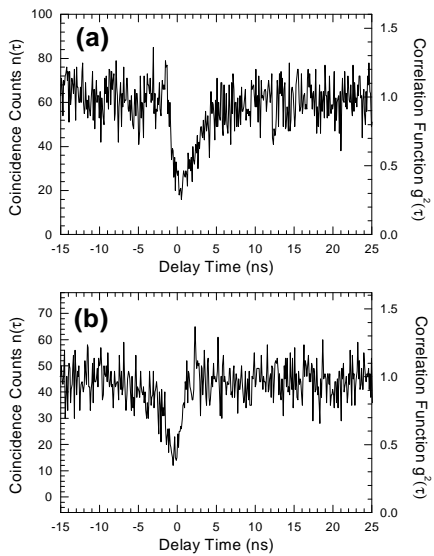}}} \caption{Cross-correlation
measurements under cw diode laser excitation at 785~nm. (a) The X2 fluorescence is sent to
the start APD while XX is sent to the stop APD. The absence of bunching demonstrates that
XX emission does not populate the X2 state. (b) The X2 emission is sent to the start APD
and the X1 emission to the stop APD. The antibunching dip shows that both transitions stem
from the same QD. The asymmetry in the dip indicates a faster recovery for the X1 state
after a X2 detection.}
\end{figure}

Having identified the two principal lines in QD spectrum, the next natural question is
whether photon correlation spectroscopy can tell us anything about the origin of the X2
emission. The cross-correlation between the X2 and XX emissions only shows antibunching
(Fig.~4(a)), indicating that while those emissions arise from the same QD, the radiative
decay of the biexciton state does not populate the X2 state. From the pump power dependent
PL spectra (Fig.~1) , it can be seen that the X2 emission has a stronger pump power
dependence than X1 but saturates earlier than the XX line. This may already suggest an
identification of X2 as a charged-exciton (trion) line. To provide further evidence, we
have carried out cross-correlation measurements between the X2 and X1 emissions
($\bar{g}^{(2)}(\tau)$) where the start and stop APDs were illuminated by the X1 and X2
lines, respectively. The resulting X1-X2 cross-correlation function (Fig.~4(b)) clearly
shows asymmetric antibunching with $\bar{g}^{(2)}(0)= 0.3$, which proves once again that
the two lines originate from the same QD. The asymmetry with fast recovery for $\tau > 0$
is expected if X2 arises from a charged exciton: the post-measurement state of
charged-exciton emission is a singly-charged QD. We would expect single-charge injection
into the QD to be much faster than triple charge injection, which in turn determines the
recovery-time for $\tau < 0$. Given that the X2 emission of this QD is stronger than we
typically see in other QDs, we could envision the presence of an acceptor or donor
impurity that increases the relative intensity of charged exciton emission. Presence of
carbon in these samples is well known, making it likely that a QD is p-doped.

Finally, it has been predicted that the radiative decay of a single QD biexciton state
will result in polarization-entangled-state generation \cite{BenPRL00}. To observe such
polarization  correlations, we have measured the polarization dependence of the X1-XX
cross-correlation. Under cw-excitation, we have seen no evidence for polarization
correlations. We believe that spin-decoherence that has been observed to occur in
nanosecond timescales for these QDs under non-resonant excitation is responsible for the
lack of polarization-correlation \cite{Paillard01}.

In summary, we have used photon auto- and cross correlation measurements to identify
dominant spectral features of a single QD, and characterize the recombination dynamics
under various excitation conditions. Given the difficulty of accurate theoretical
calculations and the richness of the QD spectra which differs significantly from one QD to
another, we believe that the techniques described here will be invaluable in understanding
individual QDs. Further experiments under different excitation conditions are needed to
understand the polarization correlations and eventually for the generation of
entangled-photon states.

This work was supported by David and Lucile Packard Fellowship and an Army Research Office
grant. C.B. acknowledges support from the Deutsche Forschungsgemeinschaft.


\end{document}